\documentclass[manuscript]{aastex}

\slugcomment{Submitted to Icarus.}

\shorttitle{Probable detection of hydrogen sulphide (H$_2$S) in Neptune's atmosphere.}
\shortauthors{Irwin et al.}

\begin{document}


\title{Probable detection of hydrogen sulphide (H$_2$S) in Neptune's atmosphere.}


\author{Patrick G. J. Irwin, Daniel Toledo and Ryan Garland}
\affil{Department of Physics, University of Oxford, Parks Rd, Oxford, OX1 3PU, UK.}
\author{Nicholas A. Teanby}
\affil{School of Earth Sciences, University of Bristol, Wills Memorial Building, Queens Road, Bristol, BS8 1RJ, UK.}
\author{Leigh N. Fletcher}
\affil{Department of Physics \& Astronomy, University of Leicester, University Road, Leicester, LE1 7RH, UK.}
\author{Glenn S. Orton}
\affil{Jet Propulsion Laboratory, California Institute of Technology, 4800 Oak Grove Drive, Pasadena, CA 91109, USA}
\author{Bruno B\'ezard}
\affil{LESIA, Observatoire de Paris, PSL Research University, CNRS, Sorbonne Universit\'e, Universit\'e Paris-Diderot, Sorbonne Paris Cit\'e, 5 place Jules Janssen, 92195 Meudon, France}
\email{patrick.irwin@physics.ox.ac.uk}



\begin{abstract}
Recent analysis of Gemini-North/NIFS H-band (1.45 -- 1.8 $\mu$m) observations of Uranus, recorded in 2010, with recently updated line data has revealed the spectral signature of hydrogen sulphide  (H$_2$S) in Uranus's atmosphere \citep{irwin18}. Here, we extend this analysis to Gemini-North/NIFS observations of Neptune recorded in 2009 and find a similar detection of H$_2$S spectral absorption features in the 1.57 -- 1.58 $\mu$m range, albeit slightly less evident, and retrieve a mole fraction of  $\sim1-3$ ppm at the cloud tops. We find a much clearer detection (and much higher retrieved column abundance above the clouds) at southern polar latitudes compared with equatorial latitudes, which suggests a higher relative humidity of H$_2$S  here. We find our retrieved H$_2$S abundances are most consistent with atmospheric models that have reduced methane abundance near Neptune's south pole, consistent with HST/STIS determinations \citep{kark11}. We also conducted a Principal Component Analysis (PCA) of the Neptune and Uranus data and found that in the 1.57 -- 1.60 $\mu$m range, some of the Empirical Orthogonal Functions (EOFs) mapped closely to physically significant quantities, with one being strongly correlated with the modelled H$_2$S signal and clearly mapping the spatial dependence of its spectral detectability. Just as for Uranus, the detection of H$_2$S at the cloud tops constrains the deep bulk sulphur/nitrogen abundance to exceed unity (i.e. $ > 4.4 - 5.0$ times the solar value) in Neptune's bulk atmosphere, provided that ammonia is not sequestered at great depths, and places a lower limit on its mole fraction below the observed cloud of (0.4 -- 1.3) $\times 10^{-5}$. The detection of gaseous H$_2$S at these pressure levels adds to the weight of evidence that the principal constituent of the 2.5 -- 3.5-bar cloud is likely to be H$_2$S ice.
\end{abstract}





\section{Introduction}

In a recent paper, we reported the detection of gaseous hydrogen sulphide (H$_2$S) in Uranus's atmosphere from Gemini-North/NIFS observations of Uranus made in 2010 \citep{irwin18}. The detection of H$_2$S in Uranus's atmosphere led us to wonder if the signature of this gas might also be detectable above the clouds in Neptune's atmosphere, in observations we obtained using the same instrument, Gemini-North/NIFS, in 2009 \citep{irwin11}. 

Like Uranus, the main clouds on Neptune are observed to have cloud tops at 2.5 -- 3.5 bar  \citep{irwin14, luszcz-cook16} and again, in the absence of any spectrally identifiable ice absorption features, authors have most commonly identified these clouds as being composed of either ammonia (NH$_3$) or hydrogen sulphide (H$_2$S) ice. This conclusion is based on the assumed presence at lower altitudes of an ammonium hydrosulphide (NH$_4$SH) cloud, which combines together in equal parts H$_2$S and NH$_3$ and leaves the more abundant molecule to condense alone at higher altitudes. Deeper in the atmosphere, \cite{depater91} analysed microwave observations of both Uranus and Neptune, recorded with the Very Large Array (VLA), and found that there was a missing component of continuum absorption that most likely arose from the pressure-broadened wings of H$_2$S lines with wavelengths of less than a few mm. They estimated the deep abundance of H$_2$S to be 10 -- $30 \times$ solar. They further concluded, building upon their previous studies \citep{depater89, depater85} that the bulk S/N ratio must exceed $5 \times$ the solar ratio for both planets, in order to limit the abundance of NH$_3$ at the observed pressure levels to be less than the detection limit of their observations. However, while H$_2$S is probably the source of the missing continuum absorption at microwave wavelengths (and is probably the main component of the 2.5--3.5-bar cloud) it has never been positively identified in Neptune's atmosphere, although its recent detection above the clouds in Uranus's atmosphere \citep{irwin18} and the many other similarities between Uranus and Neptune suggest that it is probably present.  

Following on from our Uranus analysis \citep{irwin18}, in this study we report a similar detection of gaseous H$_2$S above the cloud tops of Neptune, especially near its south pole. Its detection means that, like Uranus, Neptune may have accreted more sulphur than nitrogen during formation \citep[provided that ammonia is not partially dissolved in an ionic water ocean at great depths, e.g.][]{atreya06}, which supports it having formed further from the Sun than Jupiter and Saturn, where it was cold enough for significant abundances of H$_2$S to condense as ice. The detection of gaseous H$_2$S above Neptune's clouds also adds credibility to the likelihood that H$_2$S ice forms a significant component of the main cloud seen with a top at 2.5 -- 3.5 bar.
 
\section{Spectral Data Sources}
The main gaseous absorber in the H-band (i.e. 1.45 -- 1.8 $\mu$m) in Uranus's and Neptune's spectra is methane. The best available source of methane line data at low temperature in this range is the ``WKLMC@80K+" \citep{campargue13} line database, and its efficacy in modelling the near-IR spectra of Uranus was shown by \cite{irwin18}. Hence, we used these line data again in this study. For line shape we used a Voigt function, but with a sub-Lorentzian correction far from line centre as recommended for H$_2$-broadening conditions by \cite{hartmann02}. For  hydrogen sulphide (H$_2$S) and ammonia (NH$_3$) we used line data from HITRAN 2012 \citep{rothman13}, including their line widths and their temperature exponents, which were reported by \cite{irwin18} to be all that was available.

As described by \cite{irwin18} these line data were converted to k-distribution look-up tables, or k-tables, covering the Gemini/NIFS H-band spectral range, with 20 g-ordinates, 15 pressure values, equally spaced in log pressure between $10^{-4}$ and 10 bar, and 14 temperature values, equally spaced between 50 and 180 K. These tables were precomputed with the modelled instrument line shape of the Gemini/NIFS observations, set to be Gaussian with a full-width-half-maximum (FWHM) of 0.0003 $\mu$m, after an analysis of ARC lamp calibration spectra by \cite{irwin12b}. 

\section{Gemini/NIFS observations}

\begin{figure}
\epsscale{0.8}
\plotone{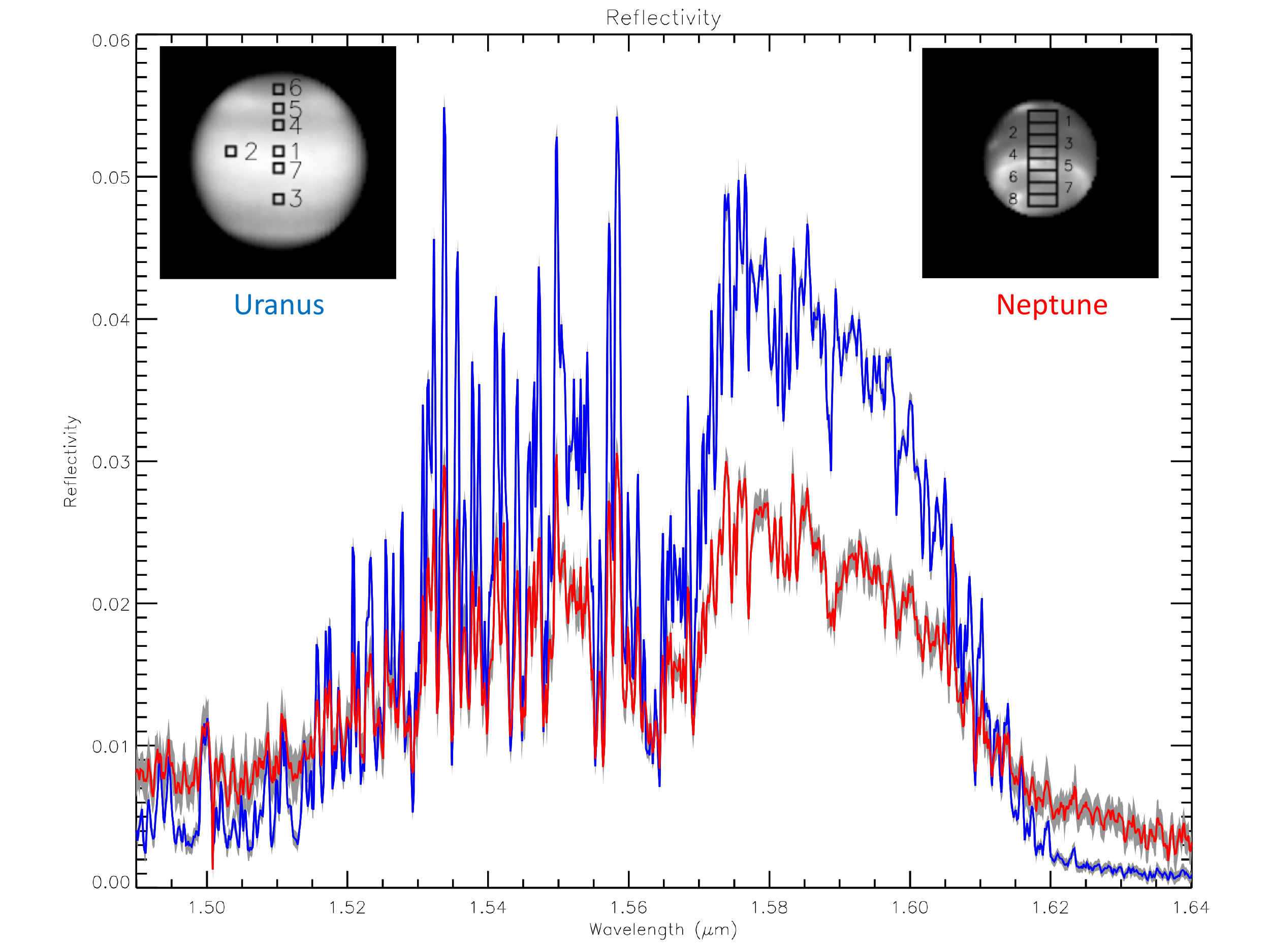}
\caption{Observed spectrum of Neptune (red) near disc centre (area `3'), with error estimates shown in grey, together with a centre-of-disc Uranus spectrum analysed by \cite{irwin18} (blue). The appearance of the planets (on the same spatial scale) near 1.58 $\mu$m is also shown for reference. The Gemini/NIFS observation of Uranus was made on 2nd November 2010 at approximately 06:00UT and the pixel areas analysed by \cite{irwin18} are indicated. The Gemini/NIFS observation of Neptune was made on 1st September 2009 at approximately 08:00UT and the eight pixel areas analysed in this paper along the central meridian are shown. We can see that overall, Uranus has higher peak reflectivity, but that Neptune shows higher reflectivity at wavelengths of strong methane absorption ($\lambda > 1.61 \mu$m and  $\lambda < 1.51 \mu$m).\label{introspec}}
\end{figure}

Observations of Neptune were made with the NIFS instrument at Gemini-North in September 2009, as reported by \cite{irwin11} and \cite{irwin14}, when Neptune presented a disc with apparent diameter of 2.35\arcsec. NIFS is an Integral Field Unit (IFU) spectrometer, which provides mapping spectrometry and returns images at 2040 wavelengths from a scene covering approximately 3\arcsec\ $\times$ 3\arcsec, with a pixel scale of 0.103\arcsec\ across slices and 0.043\arcsec\ along (sampled with a pixel size of 0.043\arcsec\ in both directions). For this study we used observations recorded on 1st September 2009 at approximately 08:00UT, which are described in detail by \cite{irwin11}.   To minimise random noise we co-added these data over a number of $13 \times 5$ pixel boxes (i.e. 0.556\arcsec\ $\times$ 0.215\arcsec, equating to a projected size at Neptune's cloud tops of $5900 \times 2300$ km), centred on the central meridian and stepped from north to south, keeping reasonably distant from the limb as shown in Fig. \ref{introspec}. This gave us eight regions to analyse in total. In Fig. \ref{introspec} we compare a typical centre-of-disc Neptune spectrum (area `3') with a typical  centre-of-disc Uranus spectrum and see that Uranus generally has higher peak reflectivity, but that Neptune shows higher reflectivity at wavelengths of strong methane absorption ($\lambda > 1.61 \mu$m and  $\lambda < 1.51 \mu$m), indicating that Neptune's atmosphere has more upper tropospheric and stratospheric haze.

We set the noise to be the standard deviation of the radiances in the averaging boxes. Ideally, we should have set the noise to be the standard error of the mean and divided these noise values by $\sqrt{13 \times 5 - 1} = 8.0$, but we found that we were unable to fit the observations to this precision; we attribute this to either deficiencies in our spectral modelling, perhaps due to residual inaccuracies in the line absorption data, or inaccuracies in our data reduction. Using the standard deviation as the noise we were able to comfortably achieve fits with $\chi^2 / n \sim 1$, which suggests that this is a more representative overall error value for our analysis. In addition, the wavelength calibration provided by the standard pipeline was found to be insufficiently accurate to match the spectral features observed, as was seen for comparable Uranus observations \citep{irwin18}. Comparison with our initial fitted spectrum led us to modify the central wavelength and wavelength step to  $\lambda_0=1.54993$ $\mu$m and $\lambda_1 = 0.00016042$ $\mu$m, respectively, which values we used in our subsequent analysis. 

\section{Vertical profiles of temperature and gaseous abundance}
\begin{figure}
\epsscale{0.8}
\plotone{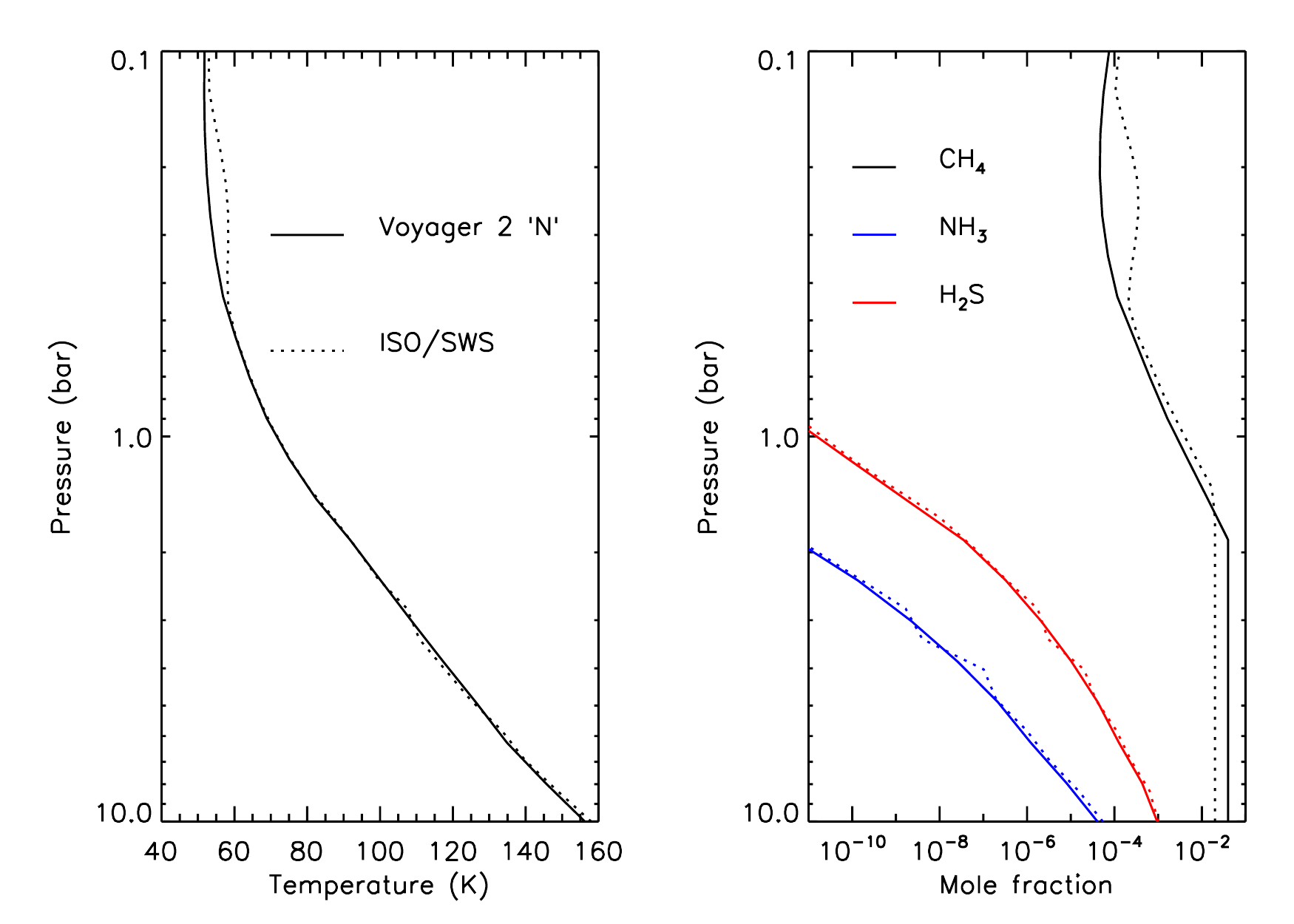}
\caption{Assumed pressure variation of temperature (left-hand panel) assumed in this study.  The reference temperature-pressure profile is based on the Voyager-2 radio-occultation `N' profile \citep{lindal92} (solid line), while the alternative profile is  the ISO temperature-pressure profile of \cite{burgdorf03} (dotted line). The right-hand panel shows the assumed profiles of condensible species. The vertical variation of the CH$_4$ abundance is as described in the text, while the abundances of NH$_3$ and H$_2$S have simply been limited by their saturation vapour pressures in both cases. We also tested a case (not shown) where the reference Voyager-2 `N' temperature-pressure profile was used, but with the deep abundance of CH$_4$ limited to 2\%.  \label{profiles}}
\end{figure}

The reference temperature and abundance profile used in this study is the same as that used by \cite{irwin14}. The temperature-pressure profile is the `N' profile determined by Voyager-2 radio-occultation measurements \citep{lindal92}, with He:H$_2$ = 0.177 (15:85), including 0.3\% mole fraction of N$_2$. The deep mole fraction of CH$_4$ was set to 4\% and at higher altitudes, where the temperature is lower, the mole fraction was limited to not exceed a relative humidity of 60\%. The mole fraction in the stratosphere was allowed to increase above the tropopause until it reached $1.5\times 10^{-3}$  \citep{lellouch10} and kept fixed at higher altitudes. To this profile we added abundance profiles of NH$_3$ and H$_2$S, assuming arbitrary `deep' mole fractions of 0.001 for both, and limited their abundance to not exceed the saturated vapour pressure in the troposphere as the temperature falls with height, and applying a `cold trap' at the tropopause to prevent the abundances increasing again in the warmer stratosphere. The abundance of H$_2$ and He at each level was then adjusted to ensure the sum of mole fractions added to 1.0 at all heights, keeping He:H$_2$ = 0.177 (15:85). These profiles are shown in Fig. \ref{profiles}.

For comparison we also performed retrievals using the temperature-pressure profile determined by \cite{burgdorf03} from Infrared Space Observatory (ISO) Short Wave Spectrometer (SWS) and Long Wave Spectrometer (LWS)  observations and ground-based mid-IR spectral observations of Neptune, assuming a deep methane mole fraction of 2\%, limited to its saturated vapour pressure curve, and `deep' NH$_3$ and H$_2$S mole fractions of 0.001.  H$_2$ and He are assumed to be present with a ratio 85:15, again ensuring the sum of mole fractions added to 1.0 at all heights. This profile was compared with the Voyager-2 radio-occultation profile and other retrievals by \cite{fletcher14}.

As a final comparison, \cite{kark11} have reported from HST/STIS observations that the `deep' methane abundance in Neptune's atmosphere decreases from $\sim 4$\% at  equatorial latitudes to $\sim 2$\% at polar latitudes. To isolate the effects of any deep variations in methane abundance we also performed retrievals with a modified version of our baseline Voyager-2 `N' profile, where the deep abundance of methane was limited to 2\%.

\section{Radiative-transfer analysis}
The vertical cloud structure was retrieved from the Gemini/NIFS observations using the NEMESIS \citep{irwin08} radiative transfer and retrieval code. NEMESIS models planetary spectra using either a line-by-line (LBL) model, or the correlated-k approximation \citep[e.g.][]{lacisoinas91}. For speed, these retrievals were conducted using the method of correlated-k, but we periodically checked our radiative transfer calculations against our LBL model to ensure they were sufficiently accurate. As with our Uranus analysis \citep{irwin18}, to model these reflected sunlight spectra, the matrix-operator multiple-scattering model of \cite{plass73} was used, with 5 zenith angles (both upwards and downwards) and the number of required azimuth components in the Fourier decomposition determined from the maximum of the reflected or incident-solar zenith angles. The collision-induced absorption of H$_2$-H$_2$ and H$_2$-He was modelled with the coefficients of  \cite{borysow92} and \cite{zhengborysow95}. Rayleigh scattering was also included for completeness, but was found to be negligible at these wavelengths. 

To analyse the measured radiance spectra within our radiative transfer model we initially used the high-resolution `CAVIAR' solar spectrum of \cite{menang13}, which we smoothed to the NIFS resolution of $\Delta\lambda = 0.0003\mu$m. However, as noted by \cite{irwin18} we found that this spectrum \citep[and others, such as those of][]{thuillier03,fiorenza05} contained spurious `Fraunhofer lines' that did not seem to correspond to features seen at these wavelengths in the Neptune spectra.  Hence, we used a smoothed version of the solar spectrum of \cite{thuillier03} in our calculations, omitting the spurious `Fraunhofer lines', which we found matched our observations much more closely. 

The observed spectra were fitted with NEMESIS using a continuous distribution of cloud particles whose opacity at 39 levels spaced  between $\sim 10$ and $\sim 0.01$ bar was retrieved. A correlation `length' of 1.5 scale heights was assumed in the \textit{a priori} covariance matrix to provide vertical smoothing. For simplicity, a single cloud particle type was assumed at all altitudes and the particles were set to have a standard Gamma size distribution \citep{hansen71} with mean radius 1.0 $\mu$m and variance 0.05, which are typical values assumed in previous analyses. Following \cite{irwin15}, the real part of the refractive index of these particles was set to 1.4 at a wavelength of 1.6 $\mu$m and NEMESIS used to retrieve the imaginary refractive index spectrum. The \textit{a priori} imaginary refractive index spectrum was sampled at every 0.05 $\mu$m between 1.4 and 1.8 $\mu$m, with a correlation length of 0.1 $\mu$m set in the \textit{a priori} covariance matrix, to ensure that retrieved spectrum varied reasonably smoothly with wavelength. At each iteration of the model, the real part of the particles' refractive index spectrum was computed using the Kramers-Kronig relation \citep{sheik05}. Self-consistent scattering properties were then calculated using Mie theory, but the Mie-calculated phase functions were approximated with combined Henyey-Greenstein functions at each wavelength to smooth over features peculiar to perfectly spherical scatterers such as the `rainbow' and `glory' as justified by \cite{irwin18}. 

\begin{figure}
\epsscale{0.8}
\plotone{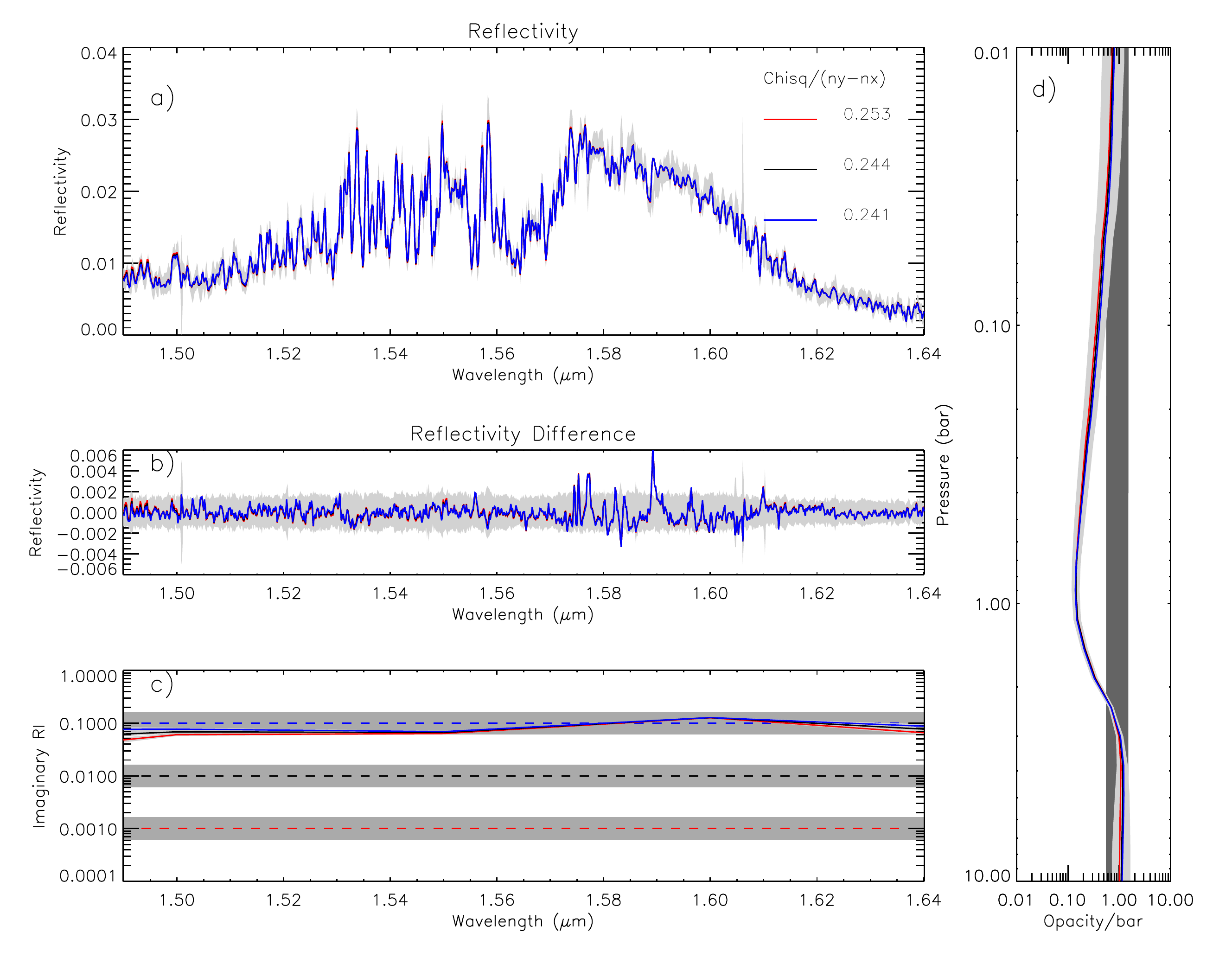}
\caption{Fit to coadded Gemini/NIFS observation of Neptune, made on 1st September 2009 at approximately 08:00UT in area `3' at 10.9$^\circ$ S, using three different assumptions for the \textit{a priori} imaginary refractive index, indicated by the coloured, dashed lines. The upper left panel compares the fitted spectra for the cases (coloured lines) with the observed spectrum and error limits (grey). The fitted $\chi^2/n$ values are indicated. The lower left panel shows the \textit{a priori} imaginary refractive indices assumed (dotted coloured lines), plus error limits (grey) and the fitted values (coloured lines) and error (dark grey). The right hand panel shows the fitted cloud opacity profiles for the three cases (opacity/bar at 1.6 $\mu$m) as coloured lines with retrieved error range as dark grey and the \textit{a priori} value and range as light grey. \label{neptunewing}}
\end{figure}

Figure \ref{neptunewing} shows our fit to our co-added Neptune spectrum in area `3' in the dark region just north of disc centre at 10.9$^\circ$S, excluding H$_2$S absorption and using three different  \textit{a priori} imaginary refractive indices of $n_i = $ 0.001, 0.01 and 0.1 ($\pm 50\%)$ at all wavelengths. Figure \ref{neptunewing} also shows our fitted cloud profiles (in units of opacity/bar at 1.6 $\mu$m) and imaginary refractive index spectra. 
Above the main retrieved cloud, with a top at 2.5 -- 3.5 bar, we find significantly more cloud opacity in the upper troposphere and lower stratosphere than we did for Uranus, consistent with previous studies, showing that the higher reflection observed at methane-absorbing wavelengths results from increased haze opacity at these altitudes. Similarly, we find no  indication of a discrete, optically significant CH$_4$ cloud at the methane condensation level of 1.5 bar, which is expected for a `background' region such as this, well away from discrete cloud features and the bright cloudy zones at 20 -- 40$^\circ$N,S. Finally, we  found a very similar dependance of the retrieved $n_i$ spectrum as for Uranus giving similar scattering properties for the particles. However, the generally higher retrieved $n_i$ values give lower single-scattering albedos of 0.6 -- 0.75. Just as for our previous analysis of Uranus's spectrum, an important consequence of the low single-scattering albedo of the retrieved particles is that solar photons are quickly absorbed as they reach the cloud tops and so we do not see significant reflection from particles existing at pressures greater than 2.5 -- 3.5 bar. Hence, although we can clearly detect the cloud top at these wavelengths, we again cannot tell where the base is and thus cannot determine whether we are seeing a vertically thin cloud based at 2.5 -- 3.5 bar, or just seeing the top of a vertically extended cloud that extends down to several bars. 

\begin{figure}
\epsscale{0.8}
\plotone{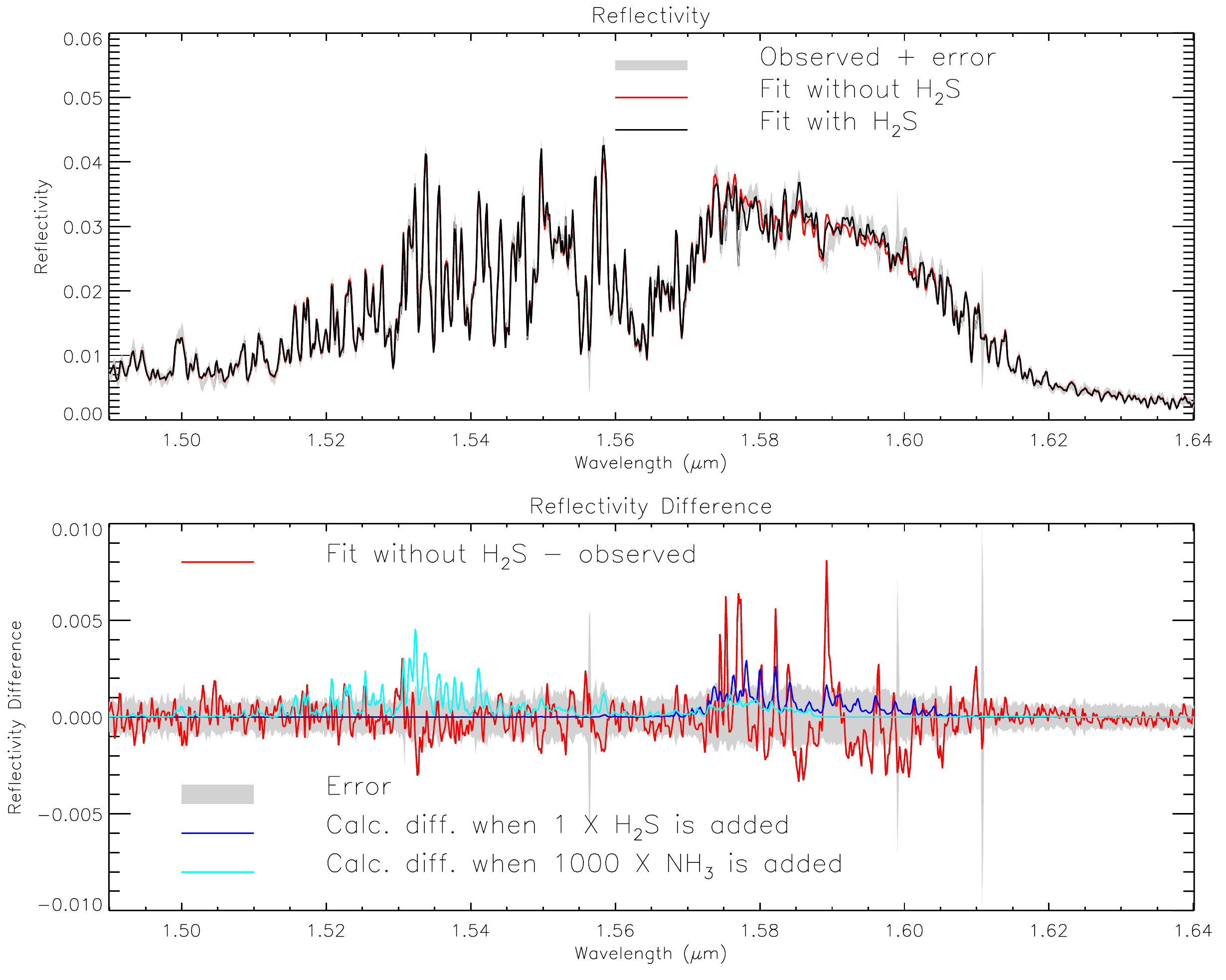}
\caption{Fit to co-added Gemini/NIFS observation of Neptune in area `7' at 58.4$^\circ$S using different assumptions. In the top plot, the observed reflectivity spectrum and estimated error is shown in grey; the fit without accounting for H$_2$S absorption is shown in red, while the fit \textbf{including} H$_2$S absorption is shown in black. The bottom plot shows the differences between the modelled and observed spectra using the same colours, with the error range shown in grey, but omits the difference plot when H$_2$S absorption is included for clarity (to allow the reader to see better the correspondence of the residual when H$_2$S is not included with the modelled difference in the spectrum when H$_2$S absorption is included/excluded). The bottom plot also shows the difference in the calculated spectra when the absorption of 100\% relative humidity (RH) of H$_2$S is included or not (blue) and when the absorption of $1000 \times$ 100\% RH of NH$_3$ is included or not (cyan). \label{neptunefit1}}
\end{figure}

\begin{figure}
\epsscale{0.8}
\plotone{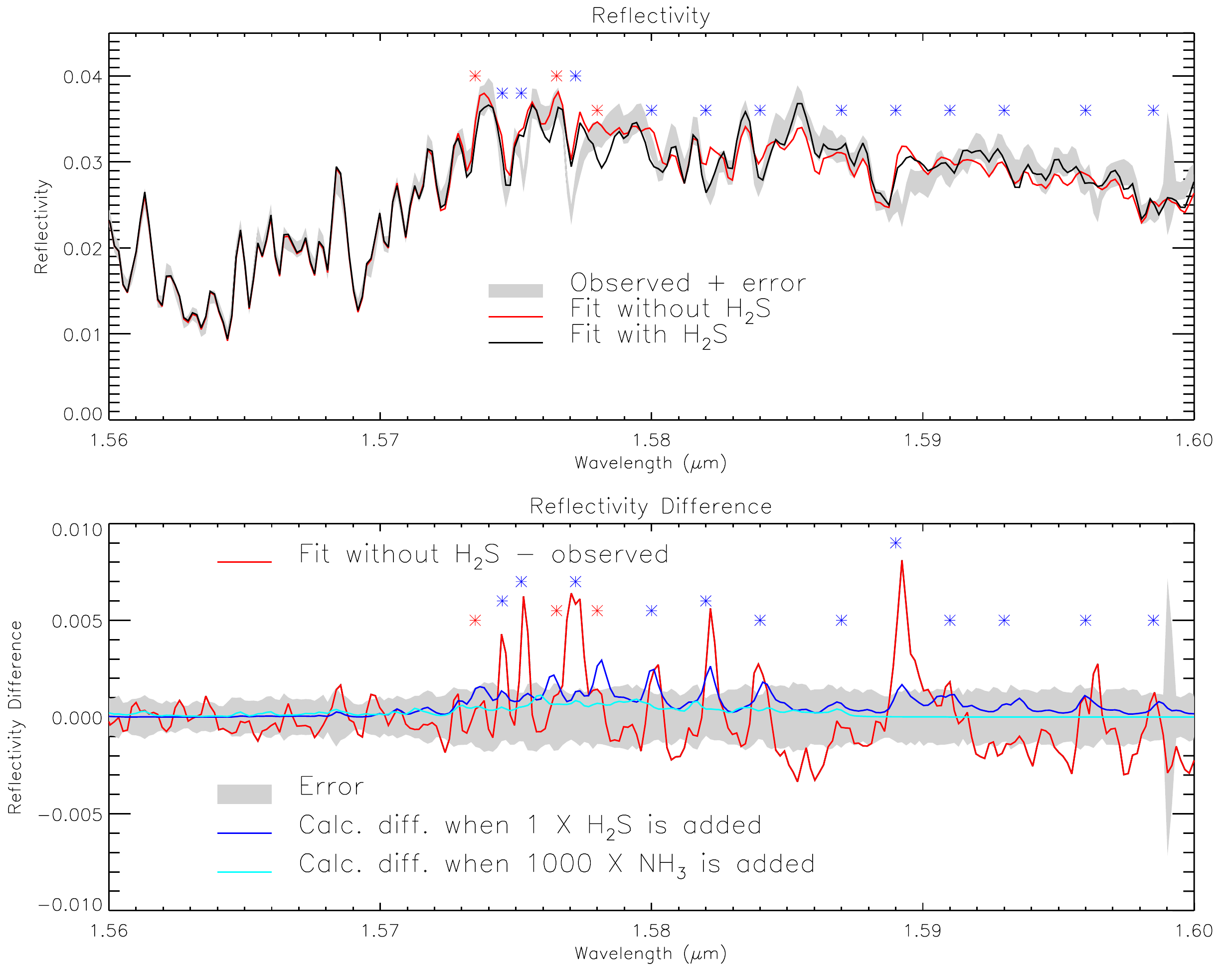}
\caption{As Fig. \ref{neptunefit1}, showing the fit to the co-added Gemini/NIFS observation of Neptune in area `7' at 58.4$^\circ$S, but expanding the 1.56 -- 1.6 $\mu$m region. Here, the features corresponding to absorption lines of H$_2$S where the fit has been significantly improved by including H$_2$S absorption are indicated by the blue asterisk symbols. Features corresponding to the few absorption lines of H$_2$S where the fit has been made worse are indicated by the red asterisk symbols. \label{neptunefit2}}
\end{figure}

\begin{figure}
\epsscale{0.8}
\plotone{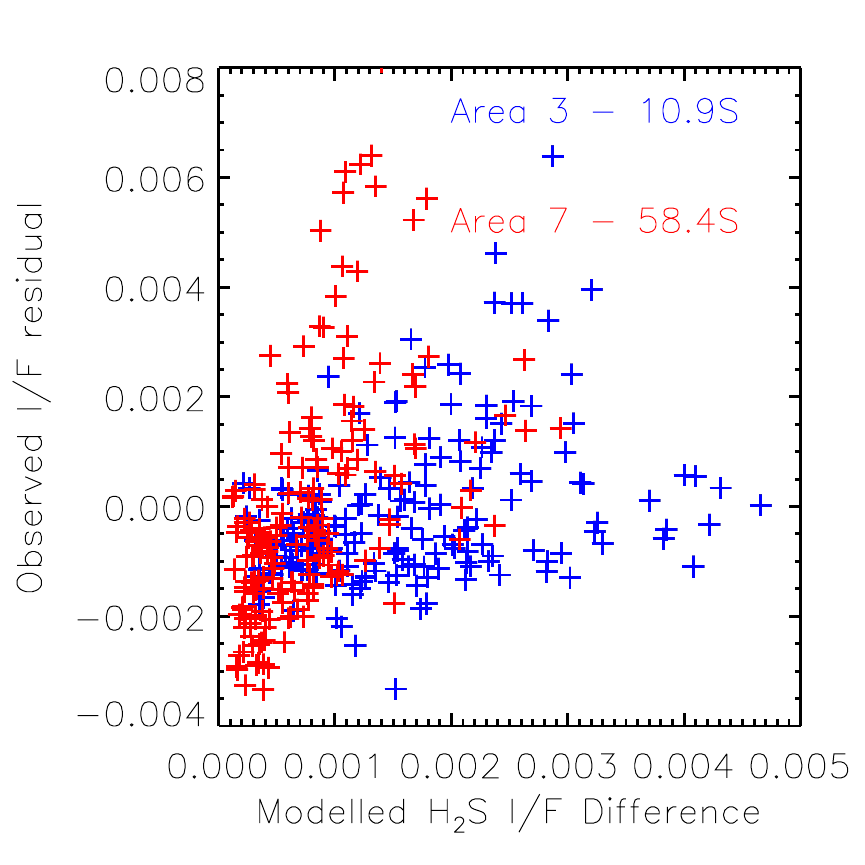}
\caption{Correlation plots of observed residual spectra when H$_2$S is excluded versus calculated difference spectra when H$_2$S absorption is included/excluded for our observations in Area `7' at 58.4$^\circ$S, showing reasonably good correlation, and Area `3' at 10.9$^\circ$S, showing weaker correlation.  \label{correlation}}
\end{figure}

\begin{figure}
\epsscale{0.8}
\plotone{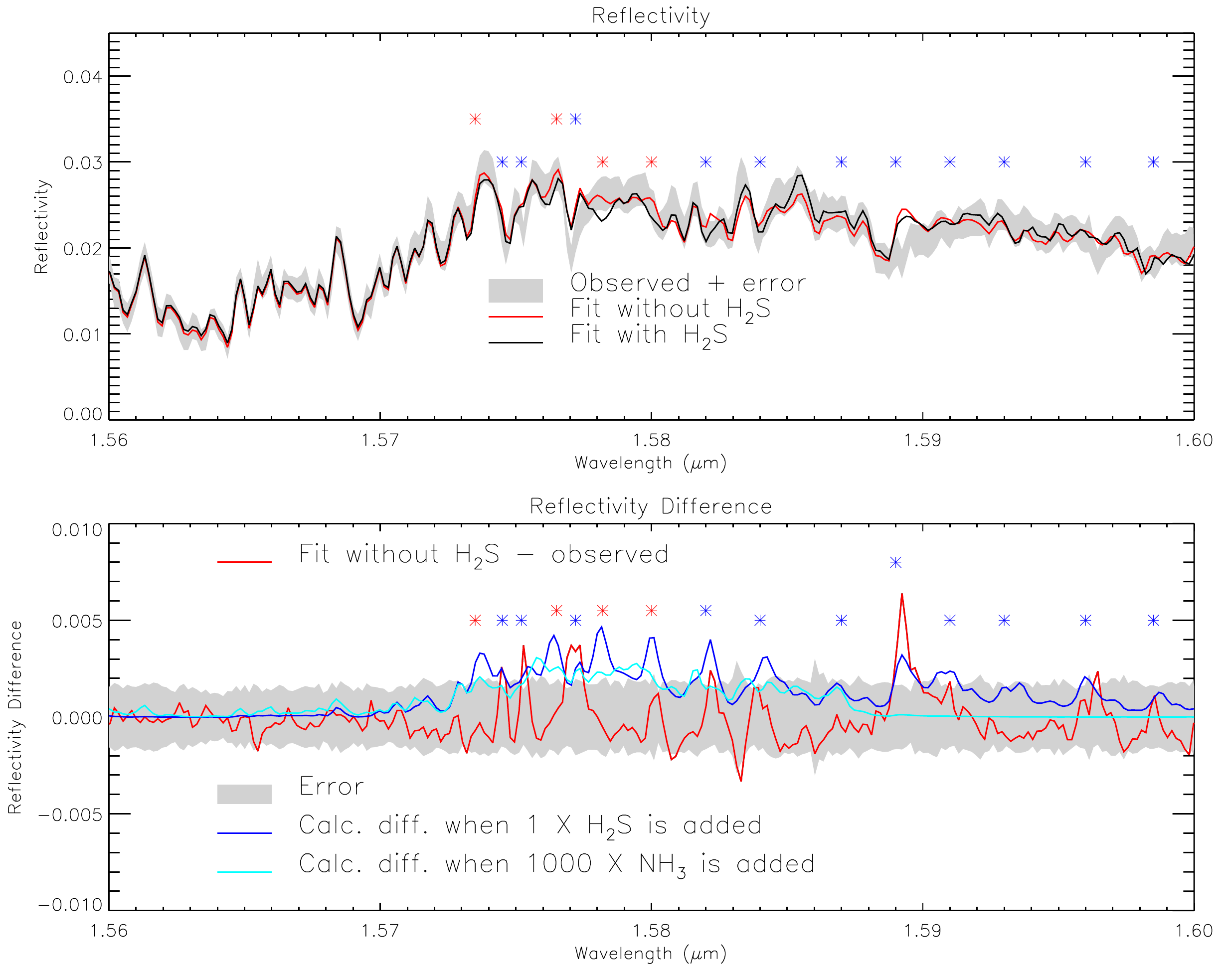}
\caption{As Fig. \ref{neptunefit2}, showing the fit in the 1.56 -- 1.6 $\mu$m region to the co-added Gemini/NIFS observation of Neptune in area `3' at 10.9$^\circ$S.  Again, features corresponding to absorption lines of H$_2$S where the fit has been significantly improved by including H$_2$S absorption are indicated by the blue asterisk symbols. Features corresponding to the few absorption lines of H$_2$S where the fit has been made worse are indicated by the red asterisk symbols. \label{neptunefit2A}}
\end{figure}

We applied our retrieval scheme, either including or excluding H$_2$S absorption, for all eight of our test areas and found the spectral signature of H$_2$S  to be more detectable near Neptune's south pole, as summarised in Table \ref{tbl-1}. Figure \ref{neptunefit1} compares our best fits to the observed co-added spectrum in area `7', centred at 58.4$^\circ$S using this model, excluding absorption by H$_2$S ($\chi^2/n$ = 1.02) and then including H$_2$S absorption ($\chi^2/n$ = 0.80). We can see that when H$_2$S absorption is not included, there is a small, but significant discrepancy between the measured and modelled spectra in the 1.575 -- 1.59 $\mu$m range, which is reduced when H$_2$S absorption is included and NEMESIS allowed to scale the H$_2$S abundance. This can be seen more clearly in Fig. \ref{neptunefit2}, where we concentrate on the 1.56 -- 1.60 $\mu$m region. We can see that when H$_2$S absorption is not included, there are several peaks in the residual reflectivity spectrum that coincide with H$_2$S  absorption features. When H$_2$S absorption is included, the fit is improved at almost all of these wavelengths, except for a few features near 1.575 $\mu$m. Note that we are generally less successful in modelling the spectrum of Neptune near 1.57 $\mu$m than for Uranus, and we will return to this point later.  We examined the correlation between the expected H$_2$S `signal' (i.e. the difference in modelled reflectivity when H$_2$S absorption is included/excluded) and the difference between the measured and fitted spectra when H$_2$S absorption is not included, in the range 1.57 -- 1.60 $\mu$m. The correlation between these two difference spectra is shown in Fig. \ref{correlation}. We found a Pearson correlation coefficient of 0.587 between these difference spectra  (indicating a reasonably strong correlation) and a Spearman's rank correlation coefficient of 0.645, with a two-sided significance value of D = $2.09 \times 10^{-23}$, equating to a 9$\sigma$-level detection. Intriguingly, this is a similar level of detection for H$_2$S as we found in our Uranus analysis, although by eye the apparent correlation between the difference spectra is less clear for Neptune than it was for Uranus. From Table \ref{tbl-1} it can be seen that we have a weaker detection of H$_2$S at more equatorial latitudes and Fig. \ref{neptunefit2A} compares the difference spectra in the 1.56 -- 1.60 $\mu$m region for the observations in area `3', centred at 10.9$^\circ$S. We can see that the residual between the measured spectrum and that fitted, omitting H$_2$S absorption, shows a poorer correlation with the modelled difference spectra when H$_2$S absorption is included/excluded. The correlation for this observation is also shown in Fig. \ref{correlation} and we find a Pearson correlation coefficient of 0.40 (indicating weaker correlation than in area `3') and a significantly lower Spearman's rank correlation coefficient of 0.3996, with a two-sided significance value of D = $1.46 \times 10^{-8}$. 

We also tested the effect on the calculated spectrum of including or excluding 100\% relative humidity of NH$_3$, but found that this was completely undetectable due to extremely low abundances of NH$_3$ at these temperatures. In case the NH$_3$ abundance in Neptune's atmosphere is in reality highly supersaturated, we also tested the effect on the calculated spectrum of supersaturating NH$_3$ by a factor of 1,000, also shown in Figs. \ref{neptunefit1} and \ref{neptunefit2}. However, we found that the absorption features of NH$_3$ do not coincide at all well with the difference spectrum, with correlation coefficients of only 0.336 (Pearson) and 0.237 (Spearman), respectively. We thus conclude, as for Uranus, that NH$_3$ is not the source of the missing absorption. 

\begin{figure}
\epsscale{0.5}
\plotone{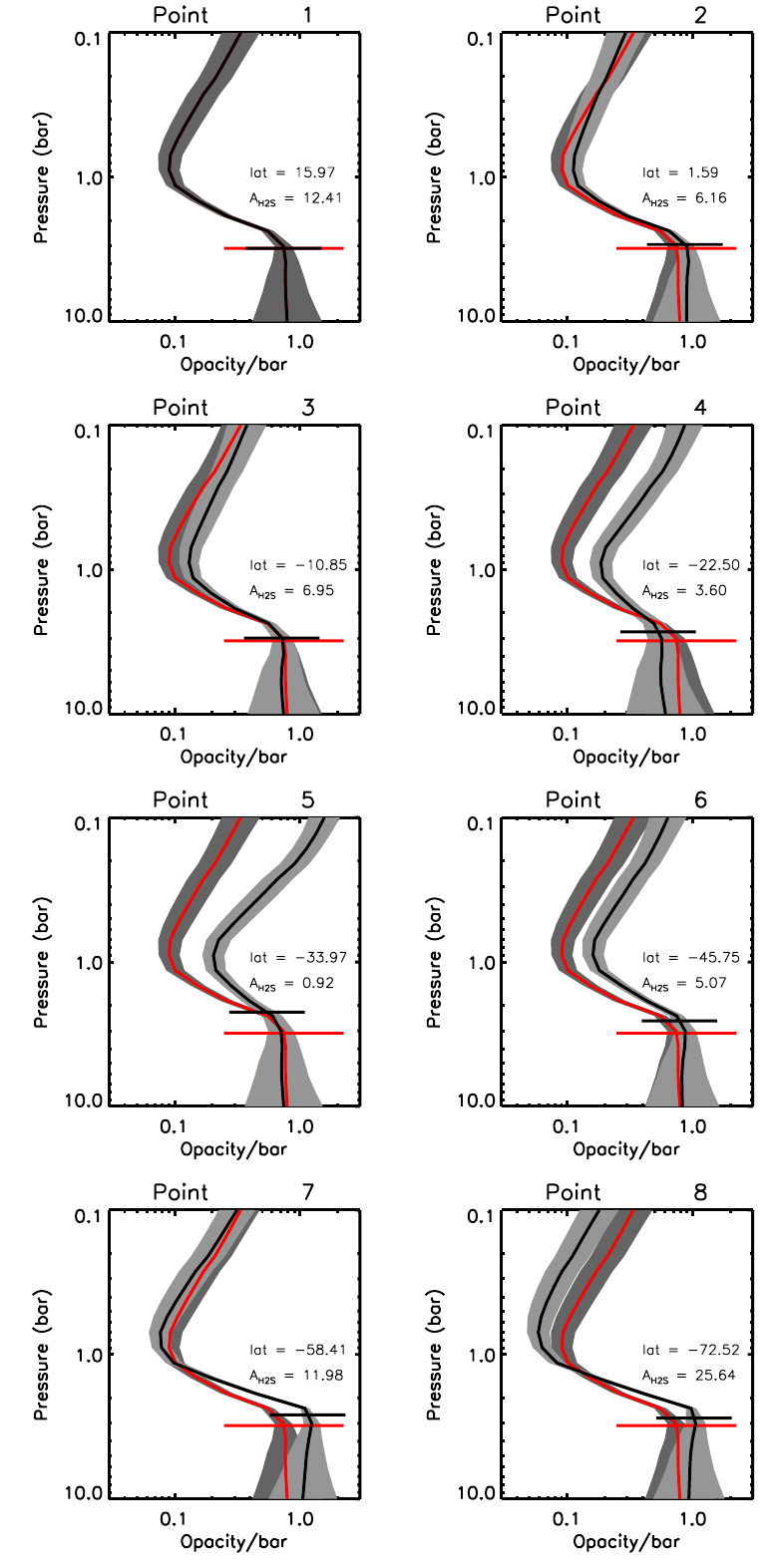}
\caption{Retrieved cloud opacity profiles in all eight test cases listed in Table \ref{tbl-1} (opacity/bar at 1.6 $\mu$m). The horizontal lines on each plot mark the pressure level where the integrated opacity to space is unity. To aid comparison, the cloud opacity profile (and cloud top pressure) retrieved for the reference pixel area `1' is shown in red for all subsequent plots. In these plots the uncertainty of the profiles are indicated in grey, where we have set the error at the ith level to be $e_i = 1/\sqrt(1/\mathbf{S}(i,i) - 1/\mathbf{S}_a(i,i))$, where \textbf{S} is the retrieved covariance matrix and \textbf{S}$_a$ is the \textit{a priori} covariance matrix. A darker grey has been used to indicate the profile error for the reference pixel area `1'.  \label{cloudsummary}}
\end{figure}

\begin{figure}
\epsscale{0.6}
\plotone{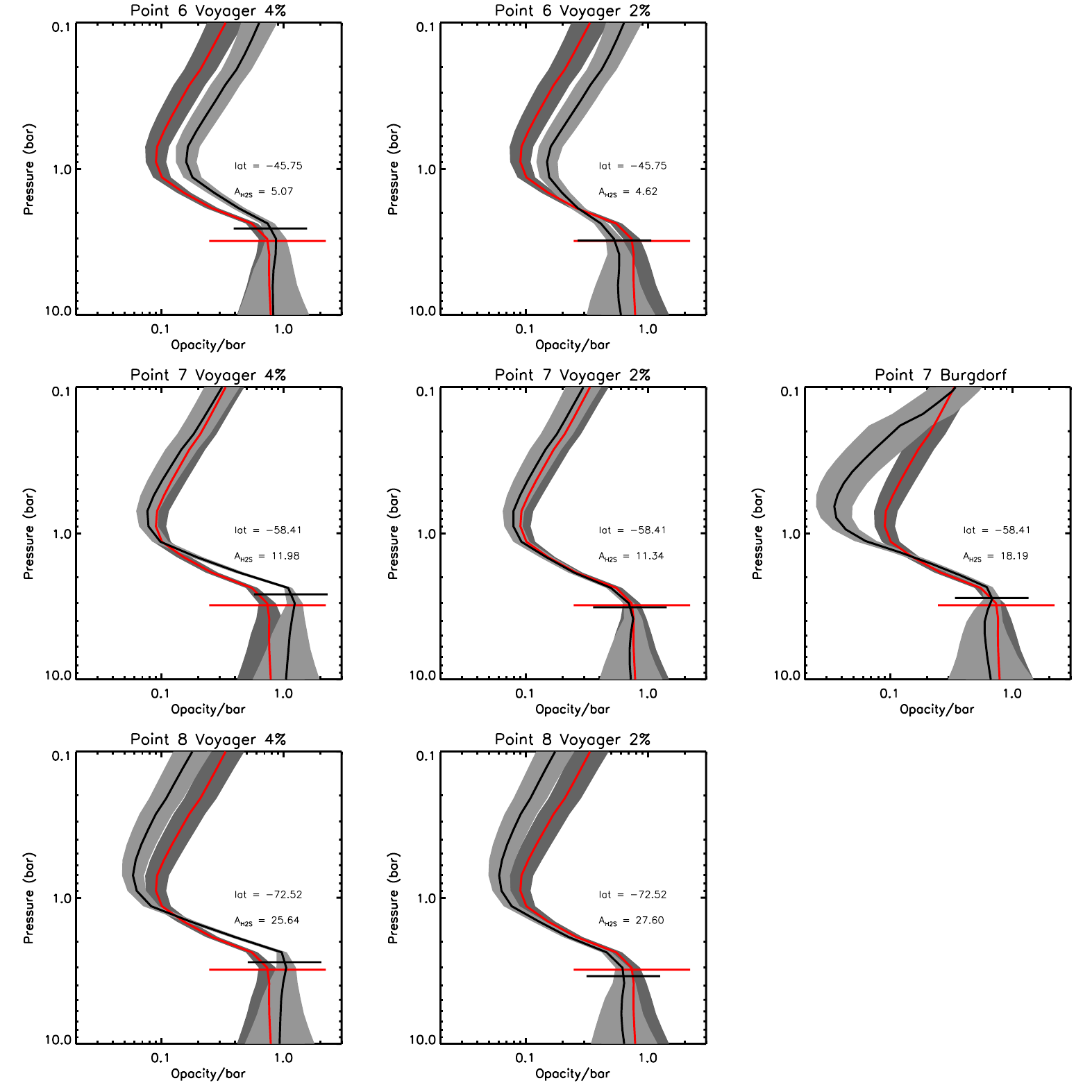}
\caption{As Fig. \ref{cloudsummary}, but comparing the retrieved cloud opacity profiles for cases 6 -- 8 listed in Table \ref{tbl-1} using different atmospheric models. The first column shows the retrievals for these areas shown in Fig. \ref{cloudsummary} using the reference Voyager 2 `N' temperature-pressure profile, with 4\% deep CH$_4$. The middle column shows our retrievals using the Voyager 2 `N' temperature-pressure profile, with 2\% deep CH$_4$ (`P'),  while the final column (for Point `7' only) shows our retrieval using the ISO temperature-pressure profile with 2\% deep CH$_4$ (`B'). As before, the horizontal lines on each plot mark the pressure level where the integrated opacity to space is unity. To aid comparison, the cloud opacity profile (and cloud top pressure) retrieved for the reference pixel area `1' using the reference temperature-pressure profile is shown in all plots in red. \label{cloudsummaryX}}
\end{figure}

\begin{figure}
\epsscale{0.6}
\plotone{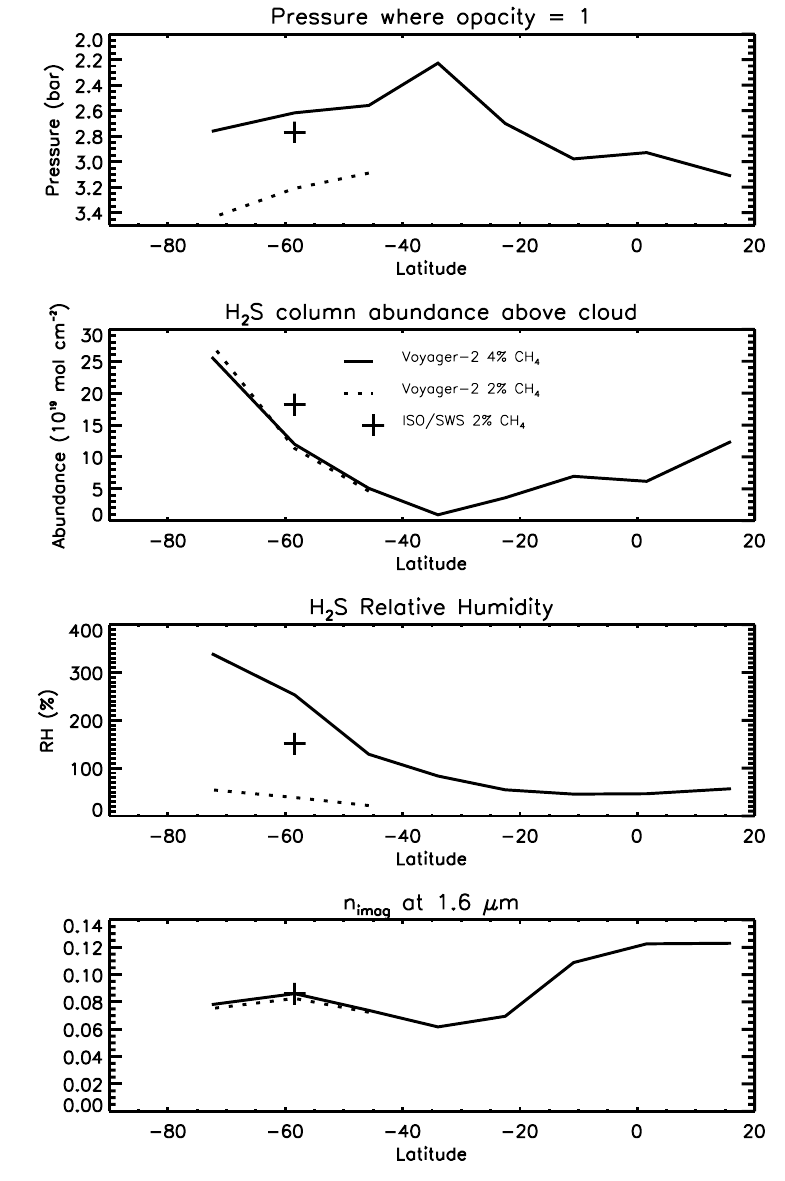}
\caption{Variation of retrieved parameters with latitude for the different atmospheric temperature-pressure profiles tested: `N' -- Voyager-2 \citep{lindal92} with 4\% deep methane; `P' -- Voyager-2 \citep{lindal92} with 2\% deep methane; and `B' -- ISO/SWS \citep{burgdorf03} with 2\% deep methane. The top panel shows the variation in retrieved cloud-top pressure $p_1$ (i.e. where the overlaying cloud opacity at 1.6 $\mu$m is unity), the upper middle panel shows the retrieved H$_2$S column abundances, while the lower middle panel shows the retrieved H$_2$S relative humidity (\%). The bottom panel shows the variation in the retrieved imaginary refractive index of the particles at 1.6 $\mu$m. The key to the line styles and symbols is shown in the upper middle panel. \label{cloudlatitude}}
\end{figure}

Our fitted cloud profiles for all eight test cases are shown in Fig. \ref{cloudsummary}. Here we can see that the cloud peaks between 2.5 and 3.5 bars for all eight locations. Furthermore, we can again see that the retrieved cloud profiles generally have enhanced cloud abundances above the main cloud deck in the 1.0 -- 0.01 bar region, compared with a similar comparison of retrievals for Uranus, shown as supplementary Fig. 11 of \cite{irwin18}. 
From Table \ref{tbl-1} we can see that the retrieved cloud-top and column abundances of H$_2$S increase towards the south pole. The corresponding retrieved relative humidity was typically $50\%$ at equatorial latitudes, but increased to values as high as $\sim 300\%$ near the south pole, which might suggest that the H$_2$S profile becomes significantly supersaturated here.  However, this conclusion may arise from inaccuracies in the assumed  temperature profile, which sets the saturated vapour pressure profile, or from inaccuracies in the assumed methane profile, which affects the retrieved cloud-top pressure and hence cloud-top temperature. To test this we repeated our retrievals for areas `6', `7' and `8' using the modified `N' Voyager-2 \citep{lindal92} temperature-pressure profile, where the deep mole fraction of CH$_4$ was reduced from from 4\% to 2\%. We also repeated the retrieval for area `7' using the vertical profile of temperature and abundance described earlier from \cite{burgdorf03}, which also has a lower deep methane abundance of 2\%. The retrieved cloud profiles are shown in Fig. \ref{cloudsummaryX} and the retrieved values summarised in Table \ref{tbl-1}. A comparison of the latitudinal dependence of the retrieved cloud-top pressure, H$_2$S column abundance, H$_2$S relative humidity, and particle imaginary refractive index at 1.6 $\mu$m for all these models is shown in Fig. \ref{cloudlatitude}. Using the original Voyager-2 `N' profile we see that, ignoring the 20 -- 40$^\circ$S cloudy zone, we retrieve significantly lower cloud-top pressures at polar latitudes than at equatorial latitudes, while using the modified Voyager-2 `N' profile, which has 2\% CH$_4$, we retrieve higher cloud-top pressures near the pole. Reducing the methane mole fraction is expected to increase the retrieved cloud-top pressure, since light needs to be reflected from deeper in the atmosphere to have the same methane column abundance, but we can see that the retrieved H$_2$S column abundances (and cloud-top mole fractions) for these two models are not significantly altered. Since the cloud-top pressure is deeper for the modified Voyager-2 profile with 2\% CH$_4$, the cloud top temperature is warmer and thus the saturated vapour pressure of H$_2$S is higher. Hence, the retrieved H$_2$S relative humidities for the modified Voyager-2 `N' profile are drastically reduced and are similar to the sub-saturated levels retrieved at equatorial latitudes using the unmodified Voyager-2 `N' profile. Similarly, using the ISO temperature-pressure profile of \cite{burgdorf03} for area `7' we find the cloud top again lies deeper in the atmosphere, as expected, although not as deep as for the modified Voyager-2 profile, which we attribute to the fact that the temperature profile is slightly different and also because this profile has more CH$_4$ at pressures less than 0.95 bar. As a result, although the retrieved H$_2$S profile has lower relative humidity, it is still supersaturated at $\sim 150\%$, compared with $\sim 250\%$ before. 

We can see from Fig. \ref{cloudlatitude} that reducing the deep CH$_4$ mole fraction from 4\% to 2\% with the Voyager-2 temperature-pressure profile leads the retrieved cloud-top pressures near Neptune's south pole to become greater than those retrieved at equatorial latitudes. If we assume that the main cloud deck is at the same pressure level at all latitudes, then we might deduce that the deep methane abundance is in reality reduced from $\sim4$\% at the equator to something more like 3\% at southern polar latitudes. This would then give similar retrieved cloud-top pressures to those found at  equatorial latitudes and would also mean that the retrieved relative humidity near the south pole would be higher than that at equatorial latitudes ($\sim$50\%), perhaps approaching 100\%. Hence, we believe these data show that the relative humidity of H$_2$S increases towards the south pole and also indirectly support the conclusion of \cite{kark11} that the deep abundance of methane reduces from 4\% near the equator to values closer to 2--3\% near the south pole in Neptune's atmosphere.

One explanation for why we retrieve higher H$_2$S relative humidities near Neptune's south pole is that the atmospheric temperatures in the 2.5 -- 3.5 bar range might possibly be warmer near the pole than they are near the equator. Since the saturated vapour pressure increases rapidly with temperature, air with a certain relative humidity in a warmer atmosphere would appear to have much higher relative humidity if analysed with a model that assumed cooler temperatures.  However, using the assumed phase curve for H$_2$S sublimation and the Voyager-2 `N' profile with 4\% methane we estimate that we would have to increase the local temperatures by almost 4K in order to reduce the retrieved relative humidity from 253\% to 100\%. \cite{fletcher14} present a reanalysis of the Voyager-2 IRIS observations of Neptune, which are sensitive to the pressure range 1.0 -- $1\times 10^{-5}$ bar and show significant variation of the retrieved temperature profile from equator to pole, with the pole and equator appearing noticeably warmer ($\sim 4$ K) than mid-latitudes at pressures of $\sim0.1$ bar (Fig. 8 of \cite{fletcher14}). However, these latitudinal variations are seen to diminish rapidly at deeper pressures, and it is thought unlikely that ice giants such as Neptune would have latitudinal temperature variations as large as 4 K at pressures greater than 1 bar due to their atmospheric circulation becoming barotropic at these pressure levels (since the circulation is dominated by convective overturning and solar heating effects are minimal). The other spectral range that allows sounding of the deep atmosphere is at radio wavelengths. \cite{depater14} show VLA radio images of Neptune at wavelengths from 0.7 to 6 cm that indicate enhanced thermal emission from the deep atmosphere near Neptune's south pole. However, these variations are interpreted as being caused by the atmosphere becoming drier at polar latitudes, allowing us to see deeper into the atmosphere, rather than due to changes in temperature. Such a conclusion is certainly supported by the latitudinal variation of methane discovered by \cite{kark11}, and supported here, but seems at odds with our conclusion that H$_2$S appears more abundant above the clouds at polar latitudes. It may be that what we detect is a cloud-top effect, rather than an increase in the H$_2$S abundance below the clouds. For example, if the clouds are `fresher' near the south pole, and so less contaminated by `sooty' photochemically-produced hydrocarbons settling down from above, then the vapour pressure of H$_2$S above the cloud particles may be higher through a process akin to Raoult's Law for the vapour pressure above liquids. This hypothesis is supported by the fact that in Fig. \ref{cloudlatitude} we can see that the retrieved imaginary refractive index of the particles is lower at polar latitudes than near the equator, indicating higher single-scattering albedoes, consistent with `fresher' particles.

\begin{figure}
\epsscale{0.8}
\plotone{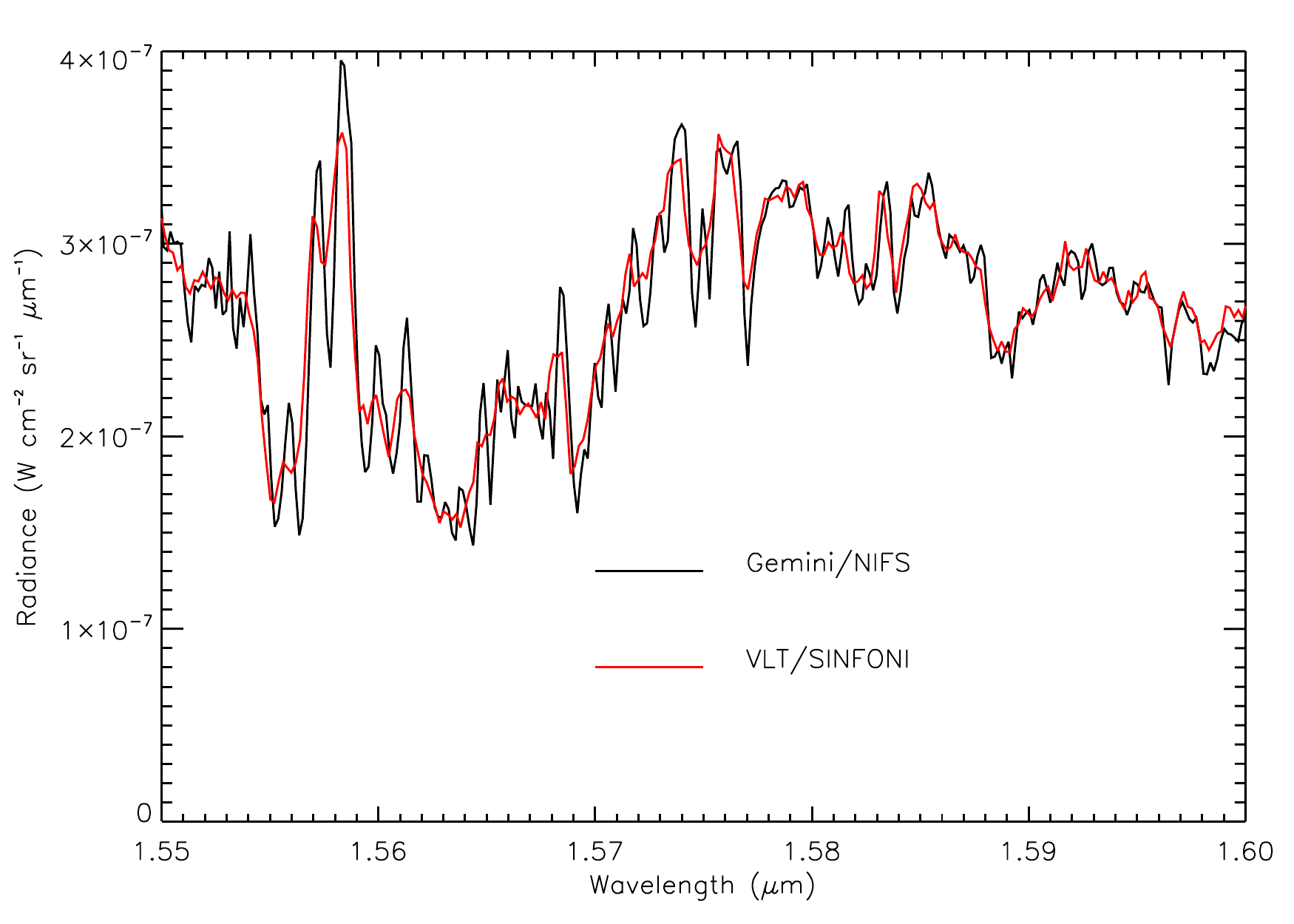}
\caption{Comparison of Gemini/NIFS spectrum in area `4' at 22.5$^\circ$S with a spectrum co-added near the disc centre and similar latitude from VLT/SINFONI observations made in 2013. \citep{irwin16}. As can be seen, the spectral features of both are well matched, although the lower spectral resolution of the SINFONI observations (R = 3000), compared with NIFS (R=5290) is apparent.\label{nifsvlt}}
\end{figure}

Comparing the measured and fitted spectra in the 1.56 -- 1.60 $\mu$m region (Fig. \ref{neptunefit2}), there are a couple of regions where our model  has difficulty in fitting the observed spectrum (which earlier meant that we had to set the noise to the standard deviation of our samples, rather than the standard error of the mean). This is most obvious near 1.59 $\mu$m, where the model seems to be missing an absorption feature, irrespective of whether H$_2$S is included or not, and an absorption feature at 1.577 $\mu$m, that is not modelled to be quite deep enough. In contrast, for the Uranus spectrum, no such discrepancies were seen \citep{irwin18}. What causes these discrepancies for Neptune, but not for Uranus is unclear, but it makes it more difficult to see the correlation between the difference spectra when H$_2$S is included/excluded. The fact that we have used the same solar spectrum for both analyses suggests that the discrepancies for Neptune are not due to mis-modelling of solar absorption lines. It is possible that the clouds themselves, which have noticeably higher retrieved imaginary refractive indices for Neptune than for Uranus (and are thus more absorbing) have additional fine structure in their true $n_i$ spectrum, not captured by the coarse resolution of our \textit{a priori} assumptions. Alternatively, it may be that our assumption of using the same particle size distribution to model the reflection at all altitudes is not appropriate for Neptune, which clearly has a higher particle density in the upper troposphere/lower stratosphere than Uranus. A further possibility is that there is some other error in the photometric correction. To test for this latter possibility we compared our Gemini-North/NIFS spectra with observations made with VLT/SINFONI in 2013 \citep{irwin16}. Figure \ref{nifsvlt} compares the spectra measured by VLT/SINFONI and Gemini/NIFS near disc centre (area `4' for Gemini/NIFS). Aside from the lower spectral resolution of the VLT/SINFONI data (R=3000, compared with R=5290 for NIFS), there is an excellent correspondence between the two sets of observations, taken four years apart from each other and calibrated independently, including in the poorly modelled regions near 1.577 and 1.59 $\mu$m.  Hence, the discrepancies between the modelled and measured spectra for Neptune seem to be real. It is clear that Neptune has more reflection from upper level hazes than Uranus and one final possibility for explaining the discrepancy is that the ``WKLMC@80K+" \citep{campargue13} line data may be less accurate at modelling methane absorption at the cooler, lower pressures of Neptune's haze layers. However, until the cause of the modelling discrepancies is isolated we must be slightly more cautious in our confidence of detection of H$_2$S in Neptune's atmosphere than we are of its detection in Uranus's atmosphere.

\section{Principal Component Analysis}
The weaker nature of the H$_2$S detection for Neptune compared with Uranus led us to explore alternative methods of detecting and mapping the distribution of H$_2$S absorption in Neptune's atmosphere and we turned to the technique of Principal Component Analysis (PCA) \citep[e.g.][]{murtagh87}, used with great success in modelling visible/near-IR Jovian spectra by \cite{dyudina01} and \cite{irwin02}. The basic idea of Principal Component Analysis is that the variance of a set of observed spectra, in this case the varying spectra observed over Neptune's disc, can be analysed into a set of Empirical Orthogonal Functions (EOFs), $E_i(\lambda)$, that form a basis from which any spectrum in the set, $y(\lambda)$, can be reconstructed as a linear combination as $y(\lambda)=\Sigma_i{\alpha_i E_i(\lambda)}$, where the coefficients, $\alpha_i$, describe the relative proportions of the different EOFs in the combined spectrum. The derived EOFs have with them an associated eigenvalue, $e_i$, and the EOFs are usually ranked in order of decreasing $e_i$. With this ordering it is  found that most of the variance can be accounted for by the first EOF (i.e. the one with the largest eigenvalue), with decreasingly significant  contributions from higher EOFs. The derived EOFs do not necessarily correspond to anything physically significant, but under certain circumstances, they can sometimes correspond to physically meaningful parameters.

\begin{figure}
\epsscale{0.3}
\plotone{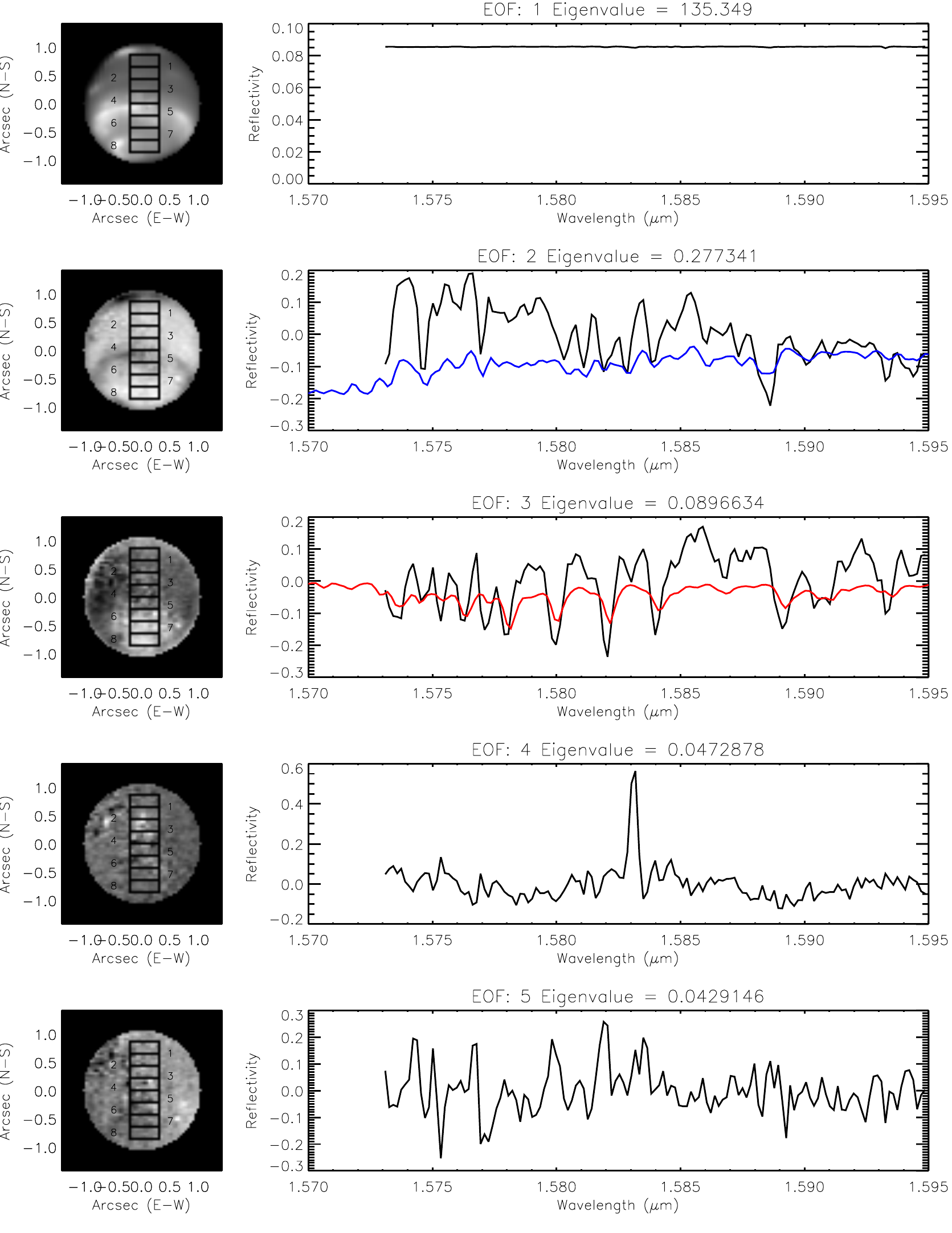}
\caption{Principal Component Analysis of Neptune observations in the spectral range 1.573 -- 1.595 $\mu$m. The right hand column shows each Empirical Orthogonal Function (EOF) derived by analysing the spectra at all locations on Neptune's disc, while the left hand column shows the relative contribution of each EOF to the observed spectrum, again at all locations on Neptune's disc. The areas chosen for our detailed retrieval analysis are also shown in the left-hand column for reference, but we must emphasise that the PCA analysis has been performed by analysing the spectra at all locations on the disc, not just the spectra in the numbered boxes. It can be seen that the eigenvalues of the EOFs fall rapidly with each EOF (indicating their becoming decreasingly significant), and we can see from the images in the left hand column that all meaningful spatial variation in the image is encapsulated in the first three EOFs. The shape of EOF 1 is almost entirely flat and this eigenfunction encapsulates the overall reflectivity as can be seen in the associated image. The spectral shape of EOF 2 is compared with the computed change in spectrum when the abundance of methane is increased (blue) and it can be seen that the associated image, to a first approximation, maps the CH$_4$ abundance above the clouds, with brighter regions having more CH$_4$ absorption. Similarly the spectral shape of EOF 3 is compared with the computed change in spectrum when the abundance of H$_2$S is increased (red) and it can be seen that the associated image maps the H$_2$S signal detectability, with brighter regions near the south pole having a higher retrieved column abundance of H$_2$S.\label{neptunepca}} 
\end{figure}

In this case, since we were interested in searching for the spectral signature of H$_2$S, whose strongest absorption lines are near 1.58 $\mu$m, we performed a principal component analysis of the observed Neptune spectra at all points on the observed disc, covering the wavelength range 1.573 -- 1.595 $\mu$m. The results are shown in Fig. \ref{neptunepca}. In this plot, the rows show the characteristics of each EOF, with the spectra showing the individual EOFs and the images showing their relative contribution to the observed spectra (i.e. the coefficients  $\alpha_i$) across the disc. The areas chosen for our detailed retrieval analysis, previously described, are also for reference. As can be seen the eigenvalues of the fitted EOFs fall rapidly and we can also see that the spatial distribution of the fitted weighting coefficients, $\alpha_i$, become more and more noisy with increasing EOF number. In fact, we found that the first three EOFs encapsulate effectively all the significant information. We can see that EOF 1 is almost entirely flat, and that its spatial map corresponds almost exactly with the I/F appearance of Neptune over these wavelengths. Hence, this EOF appears to encapsulate the overall observed mean reflectivity variation. EOF 2 contains more spectral information and we can see that its spatial distribution has low values over the main cloud belts, but high values elsewhere. We wondered whether it might be trying to encapsulate cloud height information (or equivalently methane column abundance above the clouds) and so in Fig. \ref{neptunepca} we compare the spectrum of EOF 2 with the change in the modelled reference Neptune spectrum (in this case in area `3') when we increase the methane abundance. We can see that the correlation between these two spectra is quite strong, and thus that the spatial distribution of EOF 2 can, to a first approximation, be taken as a proxy for the column abundance of methane above Neptune's clouds. EOF 3 also contains significant spectral variation, but its spatial distribution is very different from that of EOF 2, with significant contribution near Neptune's south pole, but low values everywhere else. The spectral shape of EOF 3 looks remarkably like the expected spectral signature of H$_2$S and in Fig. \ref{neptunepca} we compare it to the change in the modelled reference Neptune spectrum when we increase the H$_2$S  abundance. As can be seen, the correspondence is remarkably good. Hence, applying the PCA technique in this spectral range seems to provide a quick and effective way of mapping the detectable column abundance above the clouds of both methane and hydrogen sulphide. 

\begin{figure}
\epsscale{0.3}
\plotone{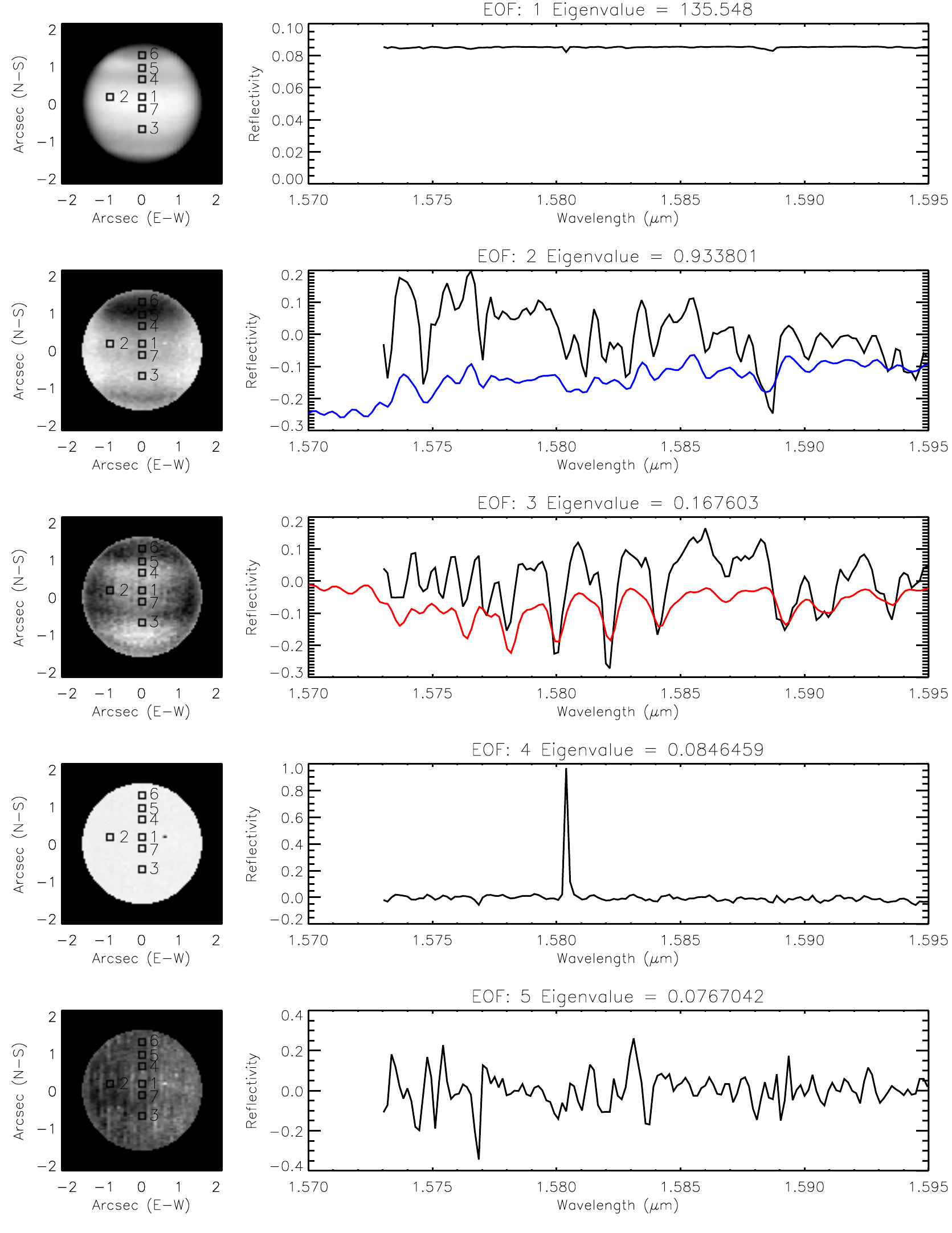}
\caption{As Fig. \ref{neptunepca}, but showing a Principal Component Analysis of Uranus observations \citep{irwin18} in the spectral range 1.573 -- 1.595 $\mu$m. Again, we see that the eigenvalues of the Empirical Orthogonal Functions (EOFs) fall rapidly with EOF number, and that all meaningful spatial variation in the image is encapsulated in the first three EOFs.  EOF 1 again encapsulates the overall reflectivity as can be seen in the associated image, EOF 2 maps the CH$_4$ abundance above the clouds, and EOF 3 maps the H$_2$S signal detectability. The blue and red lines show the change in the calculated Uranus spectrum when CH$_4$ or H$_2$S is increased.\label{uranuspca}} 
\end{figure}

As a result of this successful application of our Neptune observations, we also applied this technique to the observations of Uranus made with the same instrument on 2nd November 2010 and reported by \cite{irwin18}. The fitted EOF spectra and contribution maps are shown in Fig. \ref{uranuspca}. As can be seen the first three EOFs are almost identical to those derived for Neptune and seem to correspond once more with overall reflectivity, methane column abundance above the clouds and hydrogen sulphide abundance above the clouds. For the EOF 2 map, corresponding we believe with methane column abundance above the clouds, we see high values at low latitudes and low values over the poles, which is consistent with HST/STIS observations \citep{kark09} that the methane abundance varies with latitude in the same way. As discussed earlier, HST/STIS reports a similar latitudinal variation of methane abundance for Neptune \citep{kark11}, but the map of EOF 2 for Neptune (Fig. \ref{neptunepca}) appears different from that for Uranus with no indication of lower methane values near Neptune's south pole. Hence, EOF 2 should only be taken as a rough indicator for the methane column abundance above the clouds for Neptune and it may be that the deeper latitudinal methane abundance variation is masked: a) by the necessity of EOF 2 to describe the high clouds at 20 -- 40$^\circ$S; b) by mixing with the H$_2$S signal; or c) by some other discrepancy, related perhaps with our difficulty in modelling accurately these spectra. To test for the former possibility we re-ran the analysis on all areas south of 45$^\circ$S and between 20$^\circ$S and 20$^\circ$N (i.e. excluding the cloudy region between 20$^\circ$S and 40$^\circ$S), but found the same spatial dependence, i.e. no lowering of the methane `signal' near the pole and so no direct indication of lower CH$_4$ there. Hence, we can discount possibilty a) in our list. As for the detectability of H$_2$S,  for Uranus, the spatial variation is broadly similar to that of CH$_4$, but the highest values are seen to coincide with the dark belts just equatorward of the sub-polar bright zones, and lower values seen near the equator. Figure  \ref{uranuspca} also shows the locations of the regions analysed in detail by \cite{irwin18} and comparing the spatial distribution seen here with the retrieval results listed in Table 1 of \cite{irwin18} we see perfect correlation between the spatial distribution of the EOF 3 contribution and the retrieved column abundance of H$_2$S above the clouds, adding confidence to our conclusion that EOF 3 really does map the strength of the H$_2$S absorption signal in Uranus's atmosphere.  Because of this excellent correspondence between the map of EOF 3 and retrieved H$_2$S abundance for Uranus, where the H$_2$S signal is much stronger, we can be more confident that the same map for Neptune can also be interpreted as predominantly hydrogen sulphide column abundance above the clouds. Hence, Fig. \ref{neptunepca} shows higher detectability of the hydrogen sulphide signal over the south pole than at equatorial latitudes, just as we found in our formal retrievals. 
Finally, we note that part of the apparent difference between the EOF 3 maps for Uranus and Neptune may arise from the season. We can see that for Neptune the H$_2$S signal is strong at all latitudes near the south pole, while for Uranus, the signature seems to diminish towards the poles. However, this might just be because for Uranus we observe the polar regions at higher zenith angles and are thus unable to see as deeply. It could be that if Uranus were tipped with the south pole showing more towards the Earth we might find similarly high  H$_2$S signals at all southern polar latitudes. Similarly, the expected variation of deep CH$_4$ in Neptune's atmosphere may not be immediately obvious in EOF 2 as we observe the polar latitudes at a lower emission angles than the equatorial latitudes.

\section{Discussion}
As with our Uranus analysis \citep{irwin18} if we could be sure that the main observed cloud deck was vertically thin and composed of H$_2$S ice, then we could constrain the abundance of H$_2$S below it by equating the cloud base to the condensation level. However, as we have seen for Uranus the particles are found to be rather dark and thus we cannot tell whether we are seeing a vertically thin cloud based at 2.5 -- 3.5 bar or just the top of a vertically extended cloud that extends to several bars. Hence, once again, all we can do is derive a lower limit for the H$_2$S abundance below the clouds and above the expected NH$_4$SH cloud. In Table \ref{tbl-1} we retrieve cloud top pressures ranging from 2.6 -- 3.1 bar at equatorial latitudes. Assuming the main cloud is made of H$_2$S ice, is vertically thin and is based at 3.5 -- 4.0 bar, and that the Voyager-2 `N' temperature profile \citep{lindal92} we have assumed is correct, the saturated mole fraction of H$_2$S at the 3.5- and 4-bar levels (where the temperature is 114.0 K and 118.8 K) is estimated to be $0.6 \times 10^{-5}$ and $1.3 \times 10^{-5}$, respectively at equatorial latitudes. Alternatively, using the profile of \cite{burgdorf03}, the saturated vapour mole fraction at the 3.5- and 4-bar levels (where the temperature is 112.4 K and 117.5 K) is $0.4 \times 10^{-5}$ and $1.0 \times 10^{-5}$, respectively. Hence, we can conclude that the mole fraction of H$_2$S at pressures $>$ 3.5--4 bar, immediately below the clouds, must be $> (0.4 - 1.3) \times 10^{-5}$.  We can compare this to the expected abundances of H$_2$S  and NH$_3$ from microwave VLA studies  \citep{depater85,depater89,depater91,depater14}, summarised by \cite{irwin18}, who find that $10\times$solar H$_2$S and $2\times$solar NH$_3$ would give a residual mole fraction of H$_2$S above a deeper NH$_4$SH cloud of at least $3\times10^{-5}$, while for  $30\times$solar H$_2$S and $6\times$solar NH$_3$, the expected residual H$_2$S mole fraction increases to $9\times10^{-5}$. Our estimate seems significantly less than this, which suggests that the main cloud deck likely has a base at pressures greater than 4 bar. However, the fact that we detect H$_2$S at all at Neptune's cloud tops confirms that the deep abundance of H$_2$S must exceed that of NH$_3$ and hence that S/N $> 4.4 - 5.0 \times$ solar, depending on assumed solar composition \citep{irwin18}. We note, however, that this interpretation assumes that NH$_3$ and H$_2$S retain their deep bulk abundances at the level of the putative NH$_4$SH cloud. A number of studies \citep[e.g.][]{atreya06} note that it may be that ammonia is preferentially trapped in a supercritical water ocean (which is only predicted to exist in the ice giants, but not the gas giants) at great depth, which will lower its abundance at the NH$_4$SH level and thus leave only H$_2$S to condense at the main cloud deck we see at 2.5 -- 3.5 bar. 

\section{Conclusion}
In this study we have shown that we detect the presence of gaseous H$_2$S at the cloud tops of Neptune, and retrieve a cloud-top pressure 2.5 -- 3.5 bar, similar to the main cloud-top pressure retrieved for Uranus from similar Gemini/NIFS spectra \citep{irwin18}. However, for Neptune we find this cloud to be darker and retrieve significantly more cloud opacity in the upper troposphere/lower stratosphere. This very different vertical distribution and single-scattering albedo explains the gross observed differences between Uranus's and Neptune's spectra seen in Fig. \ref{introspec} and may also explain why the contribution of H$_2$S is more difficult to discern in Neptune's spectra since it is mixed more with reflection from aerosols near the tropopause at $\sim 0.1$ bar, where the particles are more absorbing and may have unaccounted-for absorption features, and where we are perhaps less well able to model the absorption of methane at the colder temperatures found at these pressures (temperatures of 50 -- 60K, compared with $\sim 100$K at the 2.5--3.5-bar cloud top).  However, the inclusion of H$_2$S absorption improves the fit to the Neptune spectra by a significant amount and hence we deduce that H$_2$S is present at and above the cloud tops of Neptune as we have also concluded for Uranus. 

We find that the retrieved column abundance of H$_2$S above the clouds increases as we move from equatorial to southern polar latitudes. This increase could be interpreted by Neptune's atmosphere becoming significantly supersaturated with H$_2$S at the cloud-top pressure of 2.5 to 3.5 bar at polar latitudes, but this degree of supersaturation seems unlikely at pressure levels abundantly supplied with cloud condensation nuclei. Latitudinal variations in temperature could also perhaps explain the relative humidity variations, but unrealistically large temperature variations are needed and such changes would not affect the retrieved H$_2$S column abundances which are significantly higher near Neptune's south pole. The most likely scenario is that there is higher degree of H$_2$S saturation above the clouds at southern polar latitudes, but that the need for super-saturated relative humidities is negated by a lower abundance of methane near the pole, as determined from HST/STIS observations by \cite{kark11}, which increases the retrieved cloud-top pressure, and thus temperature.

We find that a Principal Component Analysis isolates a component that matches strongly with the H$_2$S signal, and which increases from the equator to the pole as we retrieve. However, while for Uranus the H$_2$S signal and retrieved H$_2$S abundances peak at 45$^\circ$N,S and then decrease towards the poles, we find high H$_2$S column abundances in Neptune's atmospheres at all latitudes polewards of the cloudy zone at 20 -- 40 $^\circ$S. It may be that H$_2$S is just as abundant near Uranus's poles, but the current season on Uranus means that we cannot view these regions with low enough zenith angle to determine this.

As the cloud particles are retrieved to be rather dark, leading to typical single-scattering albedos of  $\varpi = 0.6 - 0.75$  and phase function asymmetries of $g \sim 0.7$, similar to Uranus, we are unable to see reflection from below the cloud tops at 2.5 -- 3.5 bar on both planets and thus cannot tell whether we might be seeing a vertically thin cloud based at 3.5 -- 4 bar, or just the top of a vertically extended cloud that extends to several bars. However, the clear detection of gaseous H$_2$S above Neptune's clouds leads us to conclude that H$_2$S ice likely forms a significant component of the main clouds at 2.5 -- 3.5 bar. Large imaginary refractive indices, such as we retrieve, are absent in the measured complex refractive index spectra of H$_2$O, CH$_4$ and NH$_3$ ices, which suggests that if Neptune's main clouds are indeed formed primarily of  H$_2$S ice, the particles may not be pure condensates, but may be heavily coated or mixed with photochemical products drizzling down from the stratosphere above, lowering their single-scattering albedos, identical to our conclusion for Uranus \citep{irwin18}.




\section{Acknowledgements}

We are grateful to the United Kingdom Science and Technology Facilities Council for funding this research and also to our support astronomers: Richard McDermid and Chad Trujillo. The Gemini Observatory is operated by the Association of Universities for Research in Astronomy, Inc., under a cooperative agreement with the NSF on behalf of the Gemini partnership: the National Science Foundation (United States), the Science and Technology Facilities Council (United Kingdom), the National Research Council (Canada), CONICYT (Chile), the Australian Research Council (Australia), Minist\'erio da Ci\^encia e Tecnologia (Brazil) and Ministerio de Ciencia, Tecnolog\'ia e Innovaci\'on Productiva (Argentina). We thank Larry Sromovsky for providing the code used to generate our Rayleigh-scattering opacities. Glenn Orton was supported by NASA funding to the Jet Propulsion Laboratory, California Institute of Technology. Leigh Fletcher was supported by a Royal Society Research Fellowship
and European Research Council Consolidator Grant (under the European Union's Horizon 2020 research and innovation programme, grant agreement No 723890) at the University of Leicester.



{\it Facilities:}   \facility{Gemini (NIFS), VLT (SINFONI)}.

\clearpage



\begin{table}[!h]
\caption{Retrieval results at all areas considered on Neptune's disc.\label{tbl-1}}
\begin {tabular}{l l l l l l l l l l l}
\hline
Area & Latitude & $p_1$ & $f_{H_2S}$  & $\chi^2/n$ & $\chi^2/n_y$ & $\Delta \chi^2$ & $x_{H_2S}$ & $A_{H_2S}$ & $R_H$ & Model \\
\hline
1 & 16.0$^\circ$N & 3.11 & $57\pm12$ & 0.48 & 0.46 & 41.0 & 1.47 & 12.4 & 16.5  &  N \\
2 & 1.59$^\circ$N & 2.92 & $46\pm9$ &  0.50 & 0.47 & 39.1 & 0.78 & 6.2 & 11.7 & N \\
3 & 10.9$^\circ$S & 2.98 & $46\pm12$ &  0.22 & 0.21 & 20.0 & 0.87 & 7.0 & 12.9 & N \\
4 & 22.5$^\circ$S & 2.70 & $55\pm19$ &  0.10 & 0.10 & 11.1 & 0.49 & 3.6 & 22.5 & N \\
5 & 34.0$^\circ$S & 2.23 & $84\pm24$ &  0.18 & 0.17 & 29.3 & 0.16 & 0.9 & 29.4 & N \\
\hline
6 & 45.8$^\circ$S & 2.56 & $129\pm28$ & 0.23 & 0.21 & 58.1   & 0.77 & 5.1 & 16.7 & N \\
6 & 45.8$^\circ$S & 3.09 & $22\pm5$ & 0.22 & 0.20 & 51.8  & 0.53 & 4.6 & 16.7 & P \\
\hline
7 & 58.4$^\circ$S & 2.62 & $253\pm29$ & 0.84 & 0.80 & 213.2 & 1.80 & 11.9& 8.4 & N \\
7 & 58.4$^\circ$S & 3.21 & $39\pm5$ & 0.80 & 0.76 & 194.4 & 1.22 & 11.3& 8.4 & P \\
7 & 58.4$^\circ$S & 2.77 & $151\pm17$  & 0.91 & 0.87 & 219.5 & 2.36 & 18.2 & 8.4 & B  \\
\hline
8 & 72.5$^\circ$S & 2.76 & $339\pm46$ & 0.55 & 0.52 & 140.0 & 3.62 & 25.6 & 7.4  & N\\
8 & 72.5$^\circ$S & 3.44 & $55\pm8$ & 0.52 & 0.49 & 129.2 & 2.73 & 27.6 & 7.4  & P\\

\hline
\end {tabular}

\small {Notes: $p_1$ is the pressure(bar) where the integrated cloud opacity (at 1.6 $\mu$m) to space is unity; $f_{H_2S}$ is the retrieved H$_2$S relative humidity (\%); $\chi^2/n$ is the reduced chi-squared statistic of the fit when H$_2$S is included, where $n = n_y - n_x = 889$; $\chi^2/n_y$ is the chi-squared statistic of the fit when H$_2$S is included, where $n_y  = 937$; $\Delta \chi^2$ is how much the $\chi^2$ of the fit reduces when H$_2$S absorption is included  -- values greater than 9 can be considered significant; $x_{H_2S}$ is the retrieved mole fraction of H$_2S$ (ppm) at $p_1$; $A_{H_2S}$ is the retrieved column amount of H$_2$S ($10^{19}$ molecule cm$^{-2}$) above $p_1$; $R_H$ is a haze `index' -- the ratio of the average radiance from 1.63 -- 1.64 $\mu$m divided by the average radiance from 1.57 --1.58 $\mu$m, expressed as \%; `Model' is atmospheric model: N = Voyager-2 `N' profile of \cite{lindal92} with 4\% deep CH$_4$, B = ISO profile of \cite{burgdorf03} with 2\% deep CH$_4$, `P' = Voyager-2 `N' profile of \cite{lindal92} with 2\% deep CH$_4$.}

\end{table}

\end{document}